# An End-to-End Framework for Dynamic Crime Profiling of Places

**Author:**
**Shailendra Kumar Gupta, Shreyanshu Shekhar, Neeraj Goel, Mukesh Saini**
2016csb1059@iitrpr.ac.in, 2016csb1060@iitrpr.ac.in, neeraj@iitrpr.ac.in, mukesh@iitrpr.ac.in

**Affiliations:**
**Department of Computer Science and Engineering, IIT Ropar, Punjab, India**

# 1. Abstract

Much effort is being made to ensure the safety of people. One of the main requirements of travellers and city administrators is to have knowledge of places that are more prone to criminal activities. To rate a place as a potential crime location, it needs the past crime history at that location. Such data is not easily available in the public domain, however, it floats around on the Internet in the form of newspaper and social media posts, in an unstructured manner though. Consequently, a large number of works are reported on extracting crime information from news articles, providing piecemeal solutions to the problem. This chapter complements these works by building an end-to-end framework for crime profiling of any given location/area. It customizes individual components of the framework and provides a Spatio-temporal integration of crime information. It develops an automated framework that crawls online news articles, analyzes them, and extracts relevant information to create a crime knowledge base that gets dynamically updated in real-time. The crime density can be easily visualized in the form of a heat map which is generated by the knowledge base. As a case study, it investigates 345448 news articles published by 6 daily English newspapers collected for approximately two years. Experimental results show that the crime profiling matches with the ratings calculated manually by various organizations.

# 2. Introduction

The information on the probability of meeting a crime at a place is essential for travellers, investors, and private companies. Police and a few government agencies have this information in bits and pieces [12]. However, it is generally not available in the public domain. Besides, the information is mostly static, and the changes are incorporated manually much later after the crime events have occurred. Consequently, to make the safety information easily available to the public, there have been a number of safety apps [19][5][22][4]. Yet, these are not much helpful as most of them depend on users' feedback/rating about a place, which may not be available in a timely manner[5]. In addition, many of us don't prefer to give feedback to such apps. This chapter proposes a framework to measure and dynamically calculate the crime score of a given location. The core idea of this approach is to exploit information present in the news articles published in different newspapers on the Internet. It extracts the date, place, and type of crime from each article, and models an aggregated crime score of a location. The aggregated score of a location depends on the number of crimes, their severity, and date. The resultant score is dynamic; that is, it changes with time. In this way, this chapter presents a real-time crime rate/score of different locations. As a case study, it focuses on some major cities of India. This knowledge will help travellers to plan their trips accordingly. Also, the crime profile can be used by the routing apps to find the safest path. This can also provide sufficient information to police forces so that they can accordingly disperse themselves in crime-prone areas. Several challenges need to be addressed to extract crime information from news articles. First, it has to separate crime-related news articles from other types. Once it has a crime article, it needs to identify the location and time of the crime. Extracting data, place, and type of crime from news articles is challenging because the information in news articles is context-specific and unstructured [11]. Further, multiple-meaning of the same word, multiple news articles on the same incident, and ambiguity between places, names, and organizations make the extraction more challenging. This chapter provides solutions to these individual problems and finally integrates these solutions to build an end-to-end framework to calculate crime density. This framework requires two databases, the first one is to store all the crawled articles (raw data) and the second one is to store all the processed data like locations, Geo-coordinates, crime score, crime type, etc. It analyzes over 345448 news articles published by six daily newspapers on the Internet over approximately two years to demonstrate the efficacy of the proposed solution, and the results are promising.

## 2.1. Background and Related Work

A lot of research attempts have been made in the past to extract relevant information from online available news articles. Arulanandam et al. [1] propose a method to detect crime location in theft-related news articles. After detecting the location using NER, the authors use conditional random fields (CRF) to classify whether the detected location is related to the crime or not. The method is evaluated only on 70 theft-related articles in New Zealand. Jayaweera et al. [7] analyze news articles published in Sri Lanka to extract crime information. The articles are classified as a crime or non-crime using an SVM classifier trained on TF-IDF representation of the articles. Location is extracted using NER. To find duplicate articles, the authors calculate the simhash value of the entities extracted from the article. The overall goal of the work is to build a database handler that supports viewing crime statistics on a map.

The crime information is also available with various government organizations. Tayal et al [21] use KNN based data mining technique to cluster data according to crime type and detect criminal attributes. Although the information received from government organizations is more reliable, it is not as up-to-date as the information extracted from the news articles. Joshi et al. [9] take a supervised approach to detect different types of crimes like thefts, homicide, and various drug offences in a crime dataset of the North Wales region, Australia. Yadav et. al. [25] also implement a system to derive state-wise crime statistics by analyzing government records of 14 years.

Researchers have explored crime analysis in various other safety-related applications as well. Sharma et al. [20] transform an article into a lower-dimensional space and apply KNN to detect crime-related articles. KNN requires a large number of labelled documents, also, it is not scalable with the number of articles. The authors employ NER based methods to detect locations in the document and calculate crime scores as the crime count. This crime score is used to find a safe path between two points. Similarly, Goel et al. [5] use the crowd-sourced safety score to find safe routes. Hassan et al. [6] find crime articles using an SVM classifier and then cluster these articles into groups according to the crime. Documents in each cluster are used to build a crime story. NER is used to find the location term. The crime locations are found by classifying the whole sentence as crime or no crime, using an SVM classifier.

Further, Rollo et al. in [18] focus on crime event visualization using news articles. They divided their whole process into 7 phases with each focusing on small tasks. Their method focused on 11 types of crimes and used around 13000 news articles. They only considered one location per article hence not counting the effect of a single article on several locations which is a commonly observed scenario for news articles. They applied their approach in the Modena province(Italy). Mukherjee et al. in [16], focus on developing a system that uses online documents for estimation and faster visualization of real-time district-wise crime scenarios of a state. Paper [3] and [15] uses the count of crime articles per location to estimate the crime rates, hence losing the crime severity information in the final results.

**Table 1.** Comparison with the related work.

| Work | Focus | Crime Article | Crime Type | Duplicate Detection | Location Detection | Location Classification | Crime Score |
|------|-------|---------------|------------|---------------------|--------------------|-------------------------|-------------|
| [1] | Theft detection | No | No | No | Yes | No | No |
| [7] | Database handler | Yes | No | Yes | Yes | No | No |
| [21] | Data mining | No | No | No | No | No | No |
| [20] | Safe navigation | Yes | No | No | No | No | Yes |
| [6] | Crime story | Yes | No | Yes | Yes | Yes | No |
| [9] | Crime Classification | No | Yes | No | No | No | No |
| [17] | Crime Mapping and Prediction | Yes | No | Yes | Yes | No | No |
| [18] | Crime Event Visualization | Yes | Yes | Yes | Yes | No | No |
| [16] | Visualization of real-time crime scenarios | Yes | No | No | Yes | No | No |
| [3] | Crime Register | Yes | No | No | Yes | No | No |
| [15] | Crime Analysis | Yes | Yes | Yes | Yes | No | No |
| This work | Crime Density | Yes | Yes | Yes | Yes | Yes | Yes |

Table 1 shows a summary of the related work. It shows that there has been only piecemeal work on crime density calculation. This chapter integrates and customizes these works to develop an end-to-end framework, which is currently running. In terms of individual components, it has improved document classification and location extraction; In terms of novelty, this is the first to consider Spatio-temporal fusion to calculate crime score.

## 2.2. Experimental Setup and Dataset

It uses six popular English news websites in India, given in Table 2. The crawler has been running daily to collect new articles from these websites since May 2018. The title, body, date, time, and URL of each article are stored in a database. Its database has around 345448 news articles (both crime and non-crime) collected over 2 years. The break-up of these articles, which are used for analysis, is shown in Table 2. The collected data was unlabelled. For verification and accuracy studies, a part of data was labelled. Around 50% of the labelled articles were crime-related. The remaining unlabelled data is used for verification of the framework. More details discussed in the Evaluation section.

**Table 2.** Popular News Websites

| Source | #Articles | Time Duration |
|---|---|---|
| TOI - Times of India | 123623 | Sep 2018 - Nov 2019 |
| Hindustan Times | 35562 | Jan 2018 - Nov 2019 |
| India Today | 33960 | Jan 2018 - Nov 2019 |
| The Hindu | 85035 | Dec 2018 - Nov 2019 |
| News18 | 36232 | Dec 2018 - Nov 2019 |
| NDTV | 30036 | March 2018 - Nov 2019 |

## 2.3. Methodology

The objective of this chapter is to create an automated framework that can estimate a reliable crime score for all possible locations by extracting information from online newspaper articles. The proposed method consists of three main steps. The first step is pre-processing. In this step, it crawls a news article from the Internet, determines whether or not it is a crime related article, and checks for existing duplicate articles. From each original crime article, in the second step, it extracts location and crime type information and builds a crime database. Finally, in the third step, it calculates the final crime score using the Spatio-temporal information. Finally, it creates a heat-map of crime score to visualize the crime density across India. Also, an interactive web interface is designed using which the real-time crime score of a location can be viewed.

## 2.4. Chapter Organization

The chapter is organized as follows. The preprocessing step is discussed in Section 2. The preprocessed articles are analyzed to build an article database in Section 3. Section 4, calculates the overall crime score in terms of the article entities in the database. The proposed work is evaluated in Section 5 and conclusions are stated in Section 6.

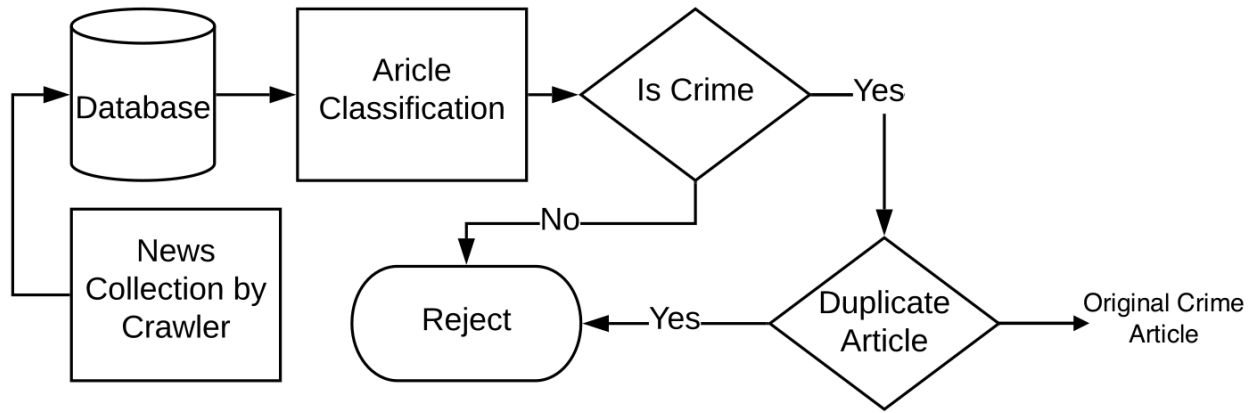

**Figure 1.** In the pre-processing step, it performs two tasks. First, it detects whether the crawled article is crime related or not. If it is a crime related article,it compares it with existing articles in the crime dataset to find whether it is an original article or a follow-up article.

# 3. Preprocessing

It starts with the collection of news articles from online news websites. Each article is classified as crime or non-crime based on the body and title. There can be multiple articles related to the same crime. Hence, crime articles are further classified as original or duplicate. Only original articles are added to the crime database. Duplicate articles are rejected. The overview of the preprocessing steps is given in Fig. 1.

## 3.1. Crime Article Detection

Intuitively, if crime-related words appear in an article, they can be classified as a crime article. However, a word can have multiple meanings depending on the context. For example, one can hit another person to commit a crime, or hit the floor to dance. Due to such context-dependent multiple meanings of the same word,  this process is very difficult. First the baseline method is explained and then an improved version being used in the system.

### 3.1.1. Baseline Method: Crime word-based Classification

In the baseline method, the crime score is mainly calculated based on the presence of the crime words. Let $W^c$ be a crime word from the initial crime word list. Not all crimes are equally severe, e.g., murder is more severe than theft [23]. These severity scores are obtained through crowd-sourcing (Section 5). Let $\psi(W^c)$ be a function that maps a crime word ($W^c$) to the corresponding severity score. The list of initial crime words is relatively small as the severity had to be crowd-source. To enrich the crime word list, it obtains synonyms of all crime words based on the similarity score. Let $W_j^c$ be a jth synonym of $W^c$. The similarity between words is calculated using the method by Wu-Palmer [24]. Let S(w1, w2) be a function that returns the similarity between w1 and w2. The severity score of a synonym word is calculated as follows:

$$\psi(w_{ij}^c) \; = \; \psi(w_i^c) \; * \; S(w_i^c, w_{ij}^c)$$

Once the severity scores are calculated for all the crime-related words (initial words and their synonyms), it looks for these words in article **A** and calculate the article severity score as follows:

$$\psi(A) \; = \; \sum_{\forall w \in A} \psi(w)$$

To classify an article as crime or no-crime, a suitable threshold is obtained through experiments. For an experimentally selected optimal threshold, it runs this algorithm on the labelled dataset, and the results are shown in Table 3. The results show that the baseline algorithm has a very high recall, but accuracy and precision are low. By analyzing the results it is found that the low accuracy was due to the ambiguity of words since one word can be used for several meanings. Further, it is observed that treating the title as a separate entity could enhance accuracy. Most related works suffer from this limitation [7].

**Table 3.** Results of Crime Classification Models

| Methods | Naive | Title Only | Body Only | Improved |
|---|---|---|---|---|
| Accuracy | 0.749 | 0.810 | 0.669 | 0.845 |

| Precision | 0.652 | 0.822 | 0.584 | 0.794 |
| Recall | 0.982 | 0.752 | 0.994 | 0.942 |

## 3.1.2. Improved Method: Ambiguity score based classification

To improve the baseline methods, it makes two important changes. First, it considers the title and body as two separate entities of a news article. Second, it introduces an ambiguity score to mitigate the effect of contextual ambiguity. The title of a news article is an abstract description of the complete article. If there is a crime word present in the title, it classifies it as a crime article. Results in Table 3 show that if it uses only the title for classification, accuracy is good, but the recall is not effective. Similarly, if it classifies based on the article's body alone, then recall is 99.6% but precision and accuracy suffer. Therefore, it treats title and body as separate features, and the overall article score is determined as a combination of both.

Contextual ambiguity is a property of a word [2]. Therefore, it counts the number of contexts in which a word can be used. Its ambiguity score is calculated as the number of crime contexts possible divided by the total number of contexts for that word. For example, "hit" can be used in two contexts, "Hit the road", or "Hit a man". Thus, the ambiguity score for "hit" is $\psi(hit)/2$. Based on the ambiguity score, it calculates the overall score of the article. In this improved approach, first classifies based on the title and body of the article. If both title and body give the same classification result, the resulting classification is given as output. However, if there is a contradiction in classification, then it uses the ambiguity score of crime words and treats both body and title as a single entity and then classifies the news article. The idea is that at least one of the classifiers should be over-confident about the class of the article in case of contradiction. The results (Table 3) show significant improvement in accuracy and precision with a slight degradation in recall in comparison to all other approaches.

**Table 4.** Confusion matrix of the duplicate detection method.

|  | Predicted | |
| --- | --- | --- |
| **Actual** | **Duplicate** | **Not duplicate** |
| **Duplicate** | 188 | 15 |
| **Not duplicate** | 90 | 537 |

## 3.2. Duplicate Article Detection

It is common for a single incident to be covered by more than one news website. On a single website also new articles are added over time to give updates on the incident. Such duplicate articles may inflate the crime score for a location if considered as a new crime. To avoid this, it needs to detect and discard all duplicate articles. If an article is a duplicate, similar words will appear in both the original and duplicate articles. Hence, it calculates the similarity between articles using Bag-of-Words (BoW) [26] representation of the news articles using TF-IDF metric [10] and applies a threshold on the similarity score to find duplicates. However, this method performs poorly because it decides the importance of words based on frequency, whereas in this case the importance of words does not depend on the frequency. Instead, article entities are more important. Main article entities are very likely to be the same in duplicate articles, e.g., both would have the same crime type, persons (victim, accused, and or investigators) and crime location.

**Table 5.** Results of duplicate detection by fixing the time span for comparison as X days, where X is 15, 30, 60, and 90 days respectively. ID refers to Article ID and Dup ID refers to respective Duplicate Article ID.

| 15 Days | | 30 Days | | 60 Days | | 90 Days | |
|---|---|---|---|---|---|---|---|
| ID | Dup ID | ID | Dup ID | ID | Dup ID | ID | Dup ID |
| 1001 | None | 1001 | 28402 | 1001 | 28402 | 1001 | 28402 |
| 1002 | 26961 | 1002 | 26961 | 1002 | 26961 | 1002 | 26961 |
| 1013 | 12948 | 1013 | 12948 | 1013 | 12948 | 1013 | 12948 |
| 1021 | 6710 | 1021 | 6710 | 1021 | 6710 | 1021 | 6710 |
| 1031 | 6663 | 1031 | 6663 | 1031 | 6663 | 1031 | 6663 |
| 1035 | 2327 | 1035 | 2327 | 1035 | 2327 | 1035 | 2327 |
| 1050 | 9503 | 1050 | 9503 | 1050 | 9503 | 1050 | 9503 |
| 1062 | None | 1062 | None | 1062 | 8586 | 1062 | 8586 |
| 1078 | None | 1078 | None | 1078 | 7852 | 1078 | 7852 |
| 1088 | None | 1088 | 7852 | 1088 | 7852 | 1088 | 7852 |

**Table 6.** time taken by the system to run a duplicate detection algorithm over 50 articles. Location means comparing only those articles which have the same crime location. Days indicates that the current article will be compared to articles that are published within X days before the current article.

| Days | Without Location(mins) | With Location(mins) |
|---|---|---|
| 15 | 67.11 | 21.74 |

| 30 | 104.99 | 28.69 |
| 60 | 146.57 | 37.18 |
| 90 | 171.50 | 44.87 |

Based on the above observations, it also matches the entities to determine the similarity of two news articles using simhash technique [7]. Hence, it has two scores, one based on the TF-IDF vectors and another based on entities. The final score is the weighted sum of both the scores.

There is a limitation of the entity score. The entities are only available in large documents. Therefore, the entity score is only used when both the documents are large. If even one document is smaller than a threshold, it uses the TF-IDF based score. Also, TF-IDF uses normalization which avoids biasing towards shorter or longer documents. The algorithm is tested by creating a new dataset. It enumerated two sets of news articles, A and B. Both sets contain 20 news articles. All articles of set A are based on the same story, whereas all articles of set B are from different stories. Finally, combining all 40 articles it randomly generated 770 pairs, which are used to produce results of this algorithm as shown in Table 4. It is able to detect duplicates with an accuracy of 94.14%.

For the framework to detect whether an incoming news article is duplicate or not, it needs to compare this news article with all other news articles present in the database. Initially, this cost of comparison will be very low due to less number of articles in the database, however, as the database increases, the number of comparisons will also shoot up, which will slow down the system. To address this problem, it uses article information to prune out some unnecessary comparisons. First, it fixes a period for comparison, i.e., articles published today are only compared with articles published in the last N months. The value of N is flexible and can be changed as per requirement. Second, there is no need to compare articles from two different locations. Hence, the overall complexity of duplicate detection will be reduced to the number of articles published at the given location for the past N months. Table 5 shows duplicate detection results for 10 articles. The result is calculated for four different periods. Table 6 shows the time taken by the system to execute the duplicate detection algorithm for 50 articles. The first column shows the time required for comparing articles with all articles published in the past 30 days from the current article's date. The second column shows the time required for comparing articles with only those articles which are published at the same location as the current article in the past 30 days. It can be observed that it saves significant time by restricting N and using the location information.

# 4. Building Crime Database

The overview of building the Crime database process is given in Fig. 2. Location, crime type, date, and named entities of each article are extracted and stored in the database. Extracting location information from an article is very tricky. In many cases, location names match precisely with a person's or organization's name. To address this challenge, the location extraction task is divided into three different parts: (1) Named entity extraction, (2) Identifying potential location entities, and (3) classifying these potential locations as crime or non-crime locations. The details of these three steps are given in the next subsections.

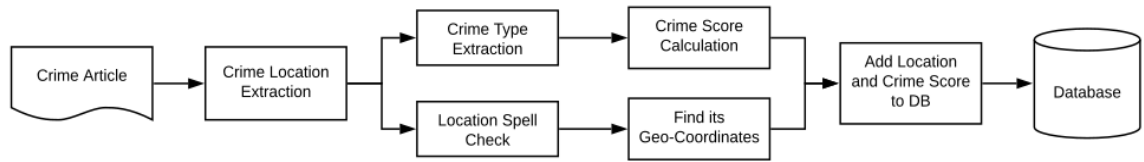

**Figure 2.** Important steps to build a crime database from the original crime articles.

**Table 7.** Accuracy improvement results for Location Separation from all entities by performing the check, presence of Common Words in entities

| Method | Without Check | With Check |
|---|---|---|
| NLTK | 52.15% | 63.21% |
| Stanford Tagger | 78.96% | 82.77% |

## 4.1. Entity Extraction

The first step is to extract all named entities in the article with tags ORGANIZATION, LOCATION/GPE, or PERSON. This chapter uses NLTK's NER [8] and Stanford NER tagger [14] for named entity tagging. NLTK's NER tagger is a supervised Maximum Entropy classifier trained on Automatic Content Extraction (ACE) data. Stanford NER tagger is computationally more expensive than NLTK's NER tagger as it uses the advanced statistical algorithms, Conditional Random Field (CRF) model. It is observed that, in both of these taggers, some names were classified as locations and vice-versa. In the next subsection, the chapter discusses an algorithm to separate names and locations.

## 4.2. Classification of entities as location or non-location

Here is one example in which both NLTK's NER tagger and Stanford NER tagger go wrong: "A man found murdered in Anand Vihar, Delhi.". In this sentence, NLTK's NER predicts Anand Vihar as location, whereas it predicts Delhi as a person entity. On the other hand, Stanford NER Tagger predicts Anand Vihar as a person entity and Delhi as a location entity. To mitigate such ambiguities, an extra country-specific step is added. If you go through the location names in India, you can observe some commonly used words in location names like Vihar, Nagar, Pradesh, Chowk, etc. Using this knowledge, a list of words, commonly used in Indian location names, is created and referred to as Common Words, and a check is performed that ensures no person entity contains any tags from this list. It is relatively less frequent that a person's name contains such tags. A tagged set of news articles is used to further analyze the efficacy of this proposed algorithm for both NLTK and Stanford NER Tagger. There were 577 locations in the tagged set of News articles. After adding this check along with NLTK's NER tagger for separating location entities from all other tagged entities an increase of 11.06% in the accuracy of NLTK's NER tagger and a 3.81% increase in the accuracy of Stanford NER Tagger is observed as shown in Table 7.

Using the ideas from the above paragraph, an algorithm is developed to classify an entity as location or non-location. First entities are extracted with their respective tags using NLTK or Stanford tagger. Next, all the entities containing any word from the Common word list are classified as locations along with the entities which are marked as LOCATION by the tagger. Finally, this combined list is returned as the set of locations from a given article. Although this solution is specific to India, changing Common words according to the applicable country is sufficient for the algorithm to work for any country.

## 4.3. Classification of extracted locations as crime or non-crime location.

The previous step only gives us a set of locations. Not all locations found in a News article are related to the crime. Following observations made from news articles helped to identify potential crime locations: (1) distance of location word from the beginning of the article, and (2) distance of location word from the crime word. It is observed that in most cases, the writer mentions the name of the crime location at the start of the article. For example, sentences "Sadar Raikot police on Wednesday booked two persons for the murder of an electrician after a fight over the phone in village Brahampura late on Tuesday evening." and "A 36-year old man was found murdered in Anand Vihar, Delhi." are the first sentences of two News articles. Similarly, it can be observed that the crime locations are mentioned closer to crime words. For example, in the sentence "Ramesh, a resident of Sultanpura, was arrested in the case of murder in Anand Vihar, Delhi.", it can be observed that the location "Anand Vihar", where the victim was murdered, is closer to crime word "murder" compared to the location "Sultanpura", which is the place of residence of accused.

The algorithm that is used to classify locations as crime or non-crime takes a set of locations, returned from the previous algorithm, and article text as input and returns a list of crime locations. First, a SentDistScore is assigned to each sentence based on its distance from the beginning of the article. Next, each location entity gets a CrimeWordDist score, the sum of the distance of all crime words from the location entity in a sentence. Here only those sentences are considered which contain that location entity. Finally, a total score is assigned to each location which is calculated as the sum of reciprocal of SentDistScore and reciprocal of CrimeWordDist. Finally, a threshold is decided empirically for separating crime locations from non-crime locations.

# 5. Crime Score Calculation

After the processing of an article is done, all the information of the article is updated into the DB. In this way, the DB will be ready with Spatio-temporal information of crime-related incidents. The goal is to fuse this extracted information and obtain an aggregated crime score. This section of the chapter explains how the extracted information is used to calculate the final crime score of a location which reveals the crime severity of that location.

## 5.1. Crime Class

Unlike many other proposed works that consider the severity of each crime type equally, the methods discussed in this chapter have divided the crimes into ten distinct crime classes and determined their severity as discussed further. These classes are shown as X-axis labels in Fig 3. Each crime class may consist of different sub-crimes. For example, 'robbery' includes 'theft', 'snatching', 'stealing' etc. Also, one word can be associated with more than one crime class. For example, death is associated with murder, suicide, accident, and terrorism. Intuitively, each class has a different crime severity. For example, murder is more severe than robbery, so its severity score should be higher. However, the judgment of the severity score is subjective. It varies from person to person. Therefore, a survey was conducted to record users' opinion about the severity of different crimes. The user is questioned, "how unsafe they would feel, on the scale of 1 to 10[1 - very safe and 10 - highly unsafe], if a particular crime out of these 10 (crime-class) has occurred at a place they are visiting". The response to the survey was post-processed and is normalized to a scale of 1 to 5. The survey was filled by 186 people. The severity score of each crime class is shown in Fig. 3. It can be observed that suicide got the minimum severity and terrorism got the maximum severity.

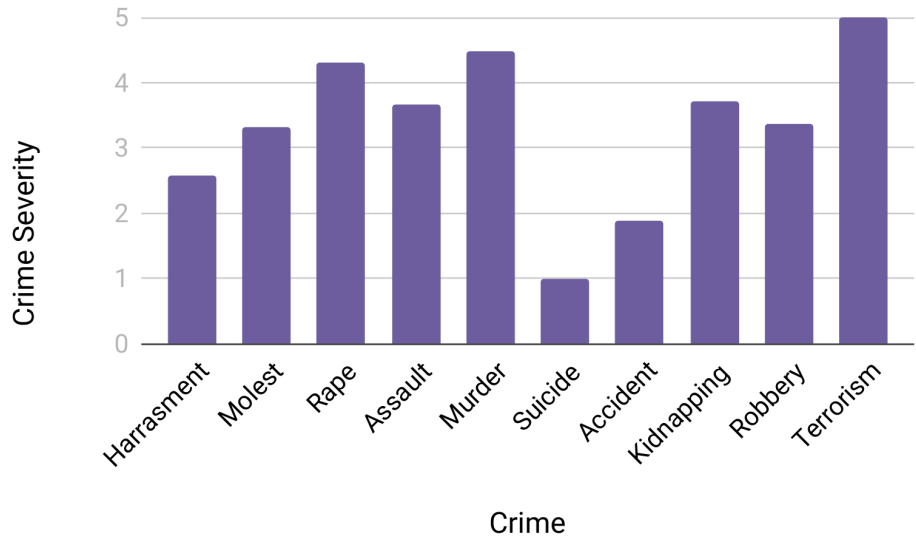

**Figure 3.** Crime Severity Score

## 5.2. Marginal Crime Score

The marginal crime score is an intermediate crime score associated with each location extracted from news articles and updated into the DB. This subsection explains how the marginal crime score is calculated. This marginal crime score of a location is updated whenever an article containing that location is processed. While processing crime articles, a crime class(mentioned in Fig. [3]) is associated with each article. Based on the crime class severity, an article score [$\gamma^{art}$] is assigned to each article. Once the article score is available, the marginal crime score of all the locations associated with this article is calculated and updated into the DB. Since the effect of a crime diminishes with time, the database is also updated with Time information. Further half-life concept is used to estimate the current marginal crime score of a location [$\gamma_c^m$] as follows:

$$\gamma_c^m = \gamma^{art} + \gamma_p^m * e^{-(\lambda * (t_c - t_p))}$$

Where

$$\lambda = \frac{ln2}{T},$$

T is half-life constant, $t_c$ is the current time, $t_p$ is the time when the marginal crime score of a location was updated last and $\gamma_p^m$ is the previous crime score; all times are in the number of days. Thus, if T is 180 days, the marginal crime score of a location will be half after every 180 days.

## 5.3. Final Crime Score

One can use this marginal score to understand the crime severity of a location or to compare the crime severity of different locations. However, this marginal score of a location is purely dependent on that location information only, whereas it is generally observed that the crime density/rate(by implication crime score) of a location does not depend only on the crime events occurring at that location but also on the crime events occurring in the neighbouring areas. Furthermore, the locations which are not present in DB(as the method has not seen any crime news article covering that location) will not be assigned any marginal crime score. To overcome the above-mentioned issues, the effect of nearby crime events is incorporated in the total crime score of a location, which is called the final crime score(or crime score) of a location. Intuitively, the impact of an incident decays with distance from the incident location [13]. This decay is assumed to be Gaussian in nature. The variance is assumed to be common for all locations. Now, whenever the crime score for a location is queried, a Gaussian is initiated centred at that location. All the locations lying within the 2.5 σ radius of queried location L are taken into account. Their marginal crime scores ($\gamma_i^{m}$) are added to estimate the final crime score of the queried location ($\gamma_L^{f}$), as follows:

$$\gamma_L^{f} = \frac{1}{\sqrt{2\pi\sigma^2}} \sum_{\forall i \in Range(2.5\sigma)} e^{-\frac{d(i,L)}{2\sigma^2}} * \gamma_i^{m}$$

where d(i, L) gives the distance between location i and L.

Before returning the crime score, it is normalized using the max marginal crime score to make sure that the returned crime score always lies between 0-1 so that any user can understand the relative safety of a location. Note that this normalized score can be greater than 1 when the queried location is having a marginal score equal to max marginal score due to the addition of scores from nearby places, in those scenarios the final crime score is rounded off to 1 to ensure that the final crime score is <= 1. Also, note that the proposed method of providing a crime score between 0-1 is more effective in comparing crime at two locations rather than giving an absolute crime index.

# 6. Auxiliary Database and APIs

While going through different newspapers it is observed that there are different spelling and even different words used by reporters for a location. These can be due to mistakes while writing or can be due to historical/cultural reasons. For example, reporters from Mumbai use Bombay, Mumbai, etc. to talk about Mumbai while writing articles. To make sure that the crime score is trustworthy even in such a diverse scenario, a separate database is created to map these words to a canonical form. Also for the cases of wrong spelling, Bing Spell Check API is used to correct the spellings of important words (like location name, entity name, etc.) before feeding them into other algorithms. In addition to this, LocationIQ API is utilized to get the Geo-coordinates of a location. The Geo-coordinates are used to calculate the distance between nearby locations when calculating the final crime score.

**Table 8.** Accuracy results for Location Extraction

| Method | Potential Locations | Crime Locations |
|---|---|---|
| NLTK | 63.21% | 60.08% |
| Stanford Tagger | 82.77% | 79.24% |

**Table 9.** Spearman's correlation coefficient ρ, showing a rank correlation between this framework and online source

| $\rho$ | Economics Times | Numbeo | Statista |
|---|---|---|---|
| **This Framework** | 0.661538 | 0.369697 | 0.785714 |

# 7. Evaluation

This section evaluates the working of the complete framework. The above framework is tested on a dataset crawled from various news websites, mentioned in Table 2. As already specified in Section 2.2, the dataset is divided into two parts. The first part (tagged manually) is used to verify various methods of the framework, and the second part is used to demonstrate the efficacy of the system. Unfortunately, there is no direct way to quantitatively measure or compare the performance of the system. To indirectly assess the quality of the final crime score, the city crime ratings are compared with the ratings published online as discussed in the latter part of this section.

The following are the details of the final framework. The best crime classification accuracy obtained after tuning all parameters is 85.4%, with 79.4% precision and 94.2% recall. More focus is given on recall to avoid missing any crime article. Crime location extraction was performed with both NLTK and Stanford tagger. Using the NLTK tagger alone, an accuracy of 75.92% is achieved in extracting all potential locations and 69.88% in separating all possible crime locations. For Stanford Tagger, an accuracy of 81.02% and 78.24% is observed respectively for both tasks. Finally, the overall framework is executed on the second part of the dataset to build a database that contains marginal crime scores and other details of crime locations.

To prove that the results are adequate, the final crime scores are cross-verified with crime reports available on some well-known websites. The results are compared with 3 online sources as listed below:

- **The Economic Times**: 7 things you didn't know about urban crime in India
- **Numbeo**: Crime Index by City
- **Statista**: Crime rate in major cities across India

The crime index/rate was collected from all these online sources and plotted into charts, as shown in Fig 4, to show the similarity between the order of their crime rating and this framework's rating. Furthermore, Spearman's Rank-Order Correlation, a non-parametric measure of rank correlation that helps in understanding the strength and direction of association between two ranked variables, is used to understand the correlation between online source crime rating and this framework's crime rating results. Some of the major locations are used for comparison which was rated by online sources as well. These major locations are ranked based on their crime score/index/rate. Next, Spearman's correlation coefficient ρ is calculated to understand the correlation between crime rating of locations from different sources with this framework results. The results are shown in Table 9. The number of samples is different in each case due to the unavailability of crime score for some locations either in online sources or in this framework. From the values, it can be seen that there is a good correlation between this framework's results and online results. Spearman's correlation coefficient value of 0.661538 is observed with the Economic Times ranking which shows a good correlation. Though the correlation is not very strong, being in line with Economic Times and other online source rankings and a good correlation with them shows that the crime score/severity estimated by this framework can be trusted.

Furthermore, the crime score results obtained are represented on an interactive map of India. To show the effectiveness of the framework at a smaller level, the heat map of New Delhi is shown in Fig. 5; the blue colour represents the colder regions, i.e. low crime occurring areas and the red colour represents the hotter regions, i.e. high crime occurring areas. This map only shows locations available in the database. To make this framework available for the end-user, a web interface is developed where the user can query the crime score of a location by providing the location name. The web interface is currently available only on the intranet of IIT Ropar. The plan is to make it publicly and freely available to the users. The source code of this framework is publicly available at *https://github.com/shreyanshu007/Crime-Analysis-BTP* git repository.

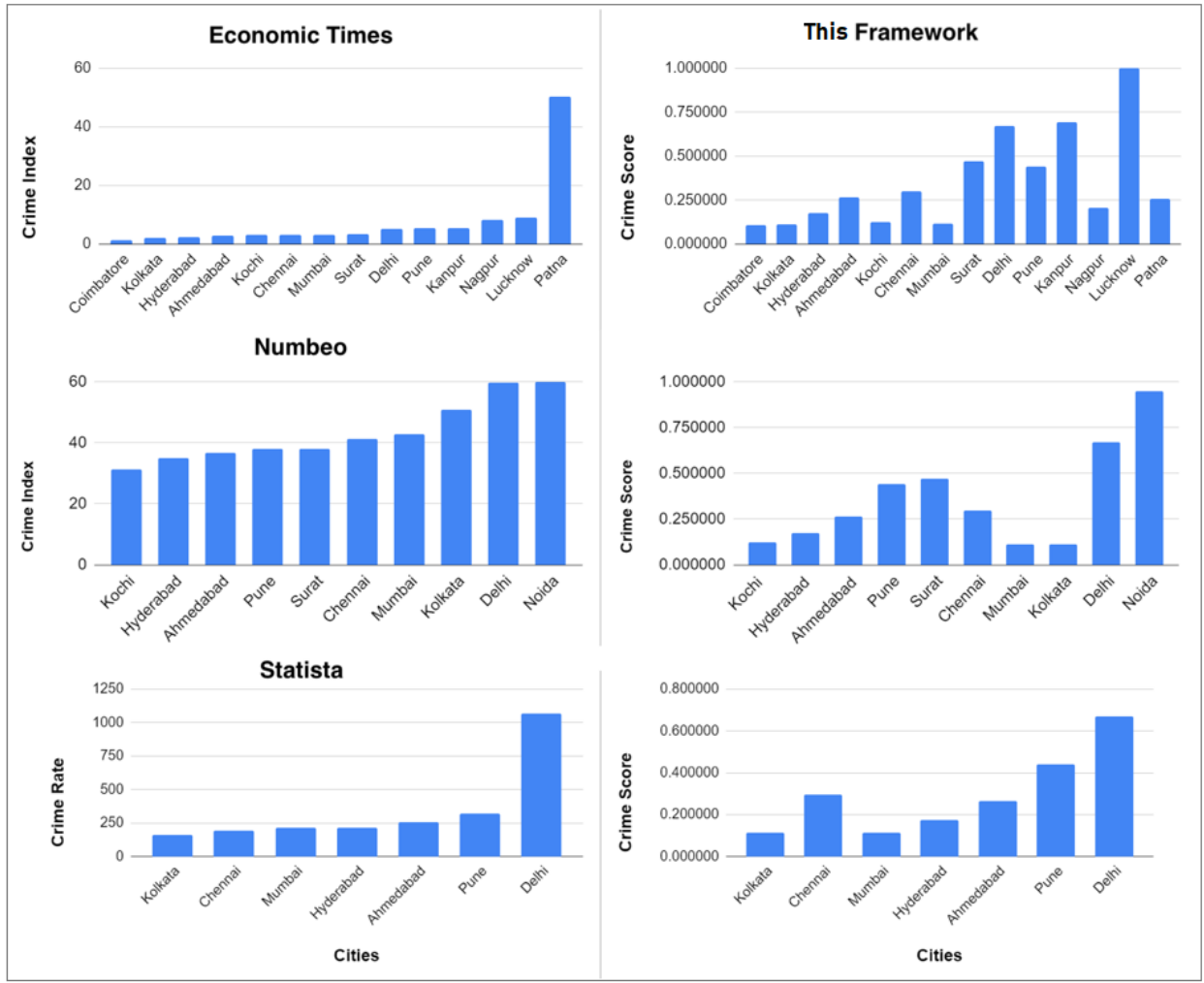

**Figure 4.** Chart showing the crime rating order of Top row: Economic times vs. this Framework; Middle row: Numbeo vs. this Framework; Bottom row: Statista vs. this Framework

# 8. Conclusion

Crime information of a location is very useful from a safety perspective. It is difficult for a normal person to get this information. This chapter of the book discusses a dynamic end-to-end framework to calculate the crime score of any given location. In the discussed framework, News articles are collected from online websites, classified into crime or non-crime, further identifying the duplicate articles and identifying crime locations, and finally, building a model for dynamic crime score for a given location. Experiments show that the crime classification accuracy is 84.5%, crime location extraction accuracy is 79.24%, and crime class identification accuracy is 76.12%. The methods can be used for crime profiling of any region as long as there is news coverage. The final crime score thus obtained is in agreement with the manual crime reports available on various online platforms. The framework is general and flexible as it can be integrated with other applications, for example, the safest route-finding application.

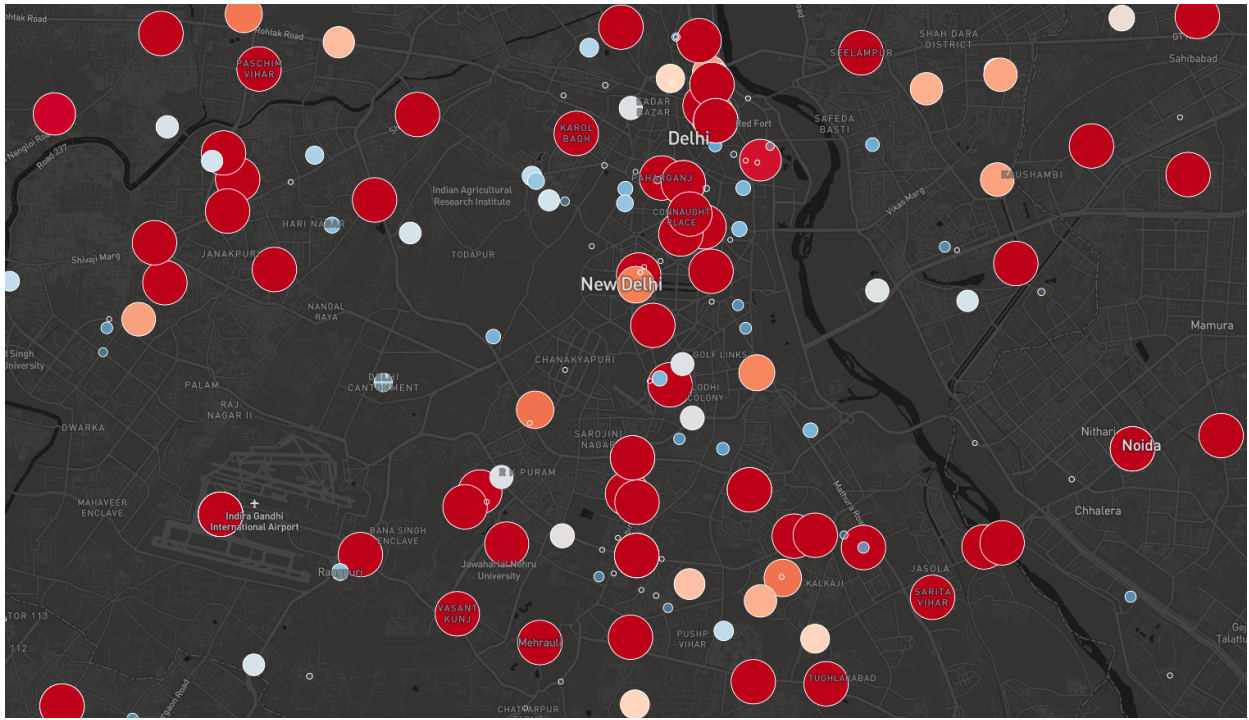

**Figure 5.** Heatmap of New Delhi and nearby locality generated using marginal crime score. The blue colour represents a low crime score while the red colour represents a high crime score.

# 9. Acknowledgments

This work was supported by the grant received from the Department of Science & Technology, Government of India, for the Technology Innovation Hub at the Indian Institute of Technology Ropar in the framework of National Mission on Interdisciplinary Cyber-Physical Systems (NM - ICPS)